\documentclass[floatfix,11pt,onecolumn,nofootinbib,amsmath,amssymb]{revtex4}
\usepackage{color}
\usepackage{epsf}
\usepackage{graphicx}  % Include figure files
\usepackage{dcolumn}   % Align table columns on decimal point
\usepackage{bm}        % bold math
\usepackage{subfigure}
\usepackage{xcolor}

\newcommand{\be}{\begin{equation}}
\newcommand{\en}{\end{equation}}
 \newcommand{\bea}{\begin{eqnarray}}
 \newcommand{\ena}{\end{eqnarray}}

\begin{document}

\title{Holographic Thermalization and Generalized Vaidya-AdS Solutions in Massive Gravity}
\author{Ya-Peng Hu$^{1,2,3}$ \footnote{Electronic address: huyp@nuaa.edu.cn}, Xiao-Xiong Zeng$^{3,4~}$\footnote{Electronic address: xxzengphysics@163.com}, Hai-Qing Zhang$^{5}$\footnote{Electronic address: H.Q.Zhang@uu.nl}}
\affiliation{ $^1$ College of Science, Nanjing University of Aeronautics and Astronautics, Nanjing 210016, China\\
$^2$ Instituut-Lorentz for Theoretical Physics, Leiden University, Niels Bohrweg 2, Leiden 2333 CA, The Netherlands \\
$^3$ State Key Laboratory of Theoretical Physics, Institute of Theoretical Physics, Chinese Academy of Sciences, Beijing, 100190, China\\
$^4$ School of Science, Chongqing Jiaotong University, Chongqing 400074, China\\
$^5$ Institute for Theoretical Physics, Center for Extreme Matter and Emergent Phenomena, Utrecht University, Princetonplein 5, 3584 CC Utrecht, The Netherlands }

%\date{ \today}

\begin{abstract}
We investigate the effect of massive graviton on the holographic thermalization process. Before doing this, we first find out the generalized Vaidya-AdS solutions in the de Rham-Gabadadze-Tolley (dRGT) massive gravity by directly solving the gravitational equations. Then, we study the thermodynamics of these Vaidya-AdS solutions by using the Minsner-Sharp energy and unified first law, which also shows that the massive gravity is in a thermodynamic equilibrium state. Moreover, we adopt the two-point correlation function at equal time to explore the thermalization process in the dual field theory, and to see how the graviton mass parameter affects this process from the viewpoint of AdS/CFT correspondence. Our results show that the graviton mass parameter will increase the holographic thermalization process.

\end{abstract}

%\pacs{04.20.-q, 04.70.-s}
%\keywords{Holographic thermalization, Vaidya solutions; Massive gravity; thermodynamics}

%\preprint{arXiv: }
 \maketitle

\section{Introduction}
Recently, the thermalization process of the quark-gluon plasma (QGP) produced in ultrarelativistic heavy-ion collisions at the Relativistic Heavy Ion Collider (RHIC) and the Large Hadron Collider (LHC) has attracted more attentions from the theoretical study~\cite{Janik:2010we,CasalderreySolana:2011us}. One of the reasons maybe come from the fact that the thermalization time of QGP predicted by the perturbation theory is longer than the experimental results \cite{Gyulassy}. Another underlying reason is that this process has been found to be strongly coupled before it hadronizes as the local temperature decreases to the deconfinement temperature~\cite{Shuryak:2008eq}. Hence, the perturbative quantum chromodynamics (QCD) breaks down in this process, while the lattice QCD is not well-suited in dealing with the real time physics or Lorentzian correlation functions~\cite{Janik:2010we,CasalderreySolana:2011us}. Therefore, other theoretical tools should be found out to describe this thermalization process.

A significant progress to deal with the strongly coupled system has been made in the last decade. The AdS/CFT correspondence has provided deep insights into the mapping between the strongly coupled field theory and the dynamics of a classical gravitational theory in the bulk, i.e. a gravitational spacetime with higher dimension~\cite{Maldacena:1997re,Gubser:1998bc,Witten:1998qj}. Moreover, the AdS/CFT correspondence has been recognized as a useful tool to deal with the strongly coupled systems~\cite{Hartnoll:2008vx,Herzog:2009xv,CasalderreySolana:2011us,Policastro:2001yc,Buchel:2003tz,Kovtun:2004de}. Therefore, for the thermalization process of strongly coupled QGP, one can also construct a proper model in the bulk gravity to investigate this process~\cite{Danielsson:1999zt}, which is usually termed as the holographic thermalization process. Indeed, until now there have been many models proposed to study the holographic thermalization process, such as~\cite{Garfinkle84, Garfnkle1202, Allais1201, Das343, Steineder, Wu1210, Gao, Buchel2013, Keranen2012, Craps1, Craps2,  Balasubramanian1, Balasubramanian2,Bai:2014tla,Chesler:2014gya,Giordano:2014kya,Roychowdhury:2016wca,Kundu:2016cgh}. Among them, a crucial model has been proposed \cite{Balasubramanian1, Balasubramanian2} by using the equal time two-point correlation function (besides space-like Wilson loop and entanglement entropy) as a thermalization probe to investigate the thermalization behavior. From the viewpoint of AdS/CFT correspondence, this is equal to probing the evolution of a shell collapsing into an AdS black brane spacetime, which can be described as a Vaidya-like solution in the bulk gravity. There have been many generalizations of this model to other cases~\cite{GS,CK, Yang1, Zeng2013, Zeng2014, Baron, Li3764, Baron1212, Arefeva, Hubeny, Arefeva6041, Balasubramanianeyal6066, Balasubramanian4, Balasubramanian3, Balasubramanian9, Cardoso, Hubeny2014, Pedraza, Pedraza2,Caceres,ZengChenLi,Alishahiha,Camilo:2014npa}.

 Many clues have been shown that the effect of massive graviton in the bulk gravity can be considered as the effect from the lattice in the dual field theory on the boundary, i.e. deducing the dissipation of momentum~\cite{Blake:2013owa,Hu:2015dnl}. Therefore, the effect from the lattice during the thermalization process of QGP may be investigated from the bulk gravity with massive graviton, i.e. massive gravity. In fact, as an extension of the Einstein's general gravity, the massive gravity is a natural generalization~\cite{Fierz:1939ix}. However, it should be noted that this generalization is difficult, since the massive gravity usually has the instability problem of the Boulware-Deser ghost~\cite{Boulware:1973my}. For more details, one can refer to the reviews of massive gravity~\cite{Hinterbichler:2011tt,deRham:2014zqa}. Recently, the so called de Rham-Gabadadze-Tolley (dRGT) massive gravity has been proposed \cite{deRham:2010ik, deRham:2010kj,deRham:2014zqa}, which is a nonlinear massive gravity theory and has been found to be ghost-free \cite{Hassan:2011hr,Hassan:2011tf}. Note that, the dRGT massive gravity is thought to be the only healthy theory in Poincar\'e invariant setups, and there have been many researches in dRGT massive gravity~\cite{Vegh:2013sk,Adams:2014vza,Cai:2014znn,Hu:2015xva,Xu:2015rfa,Hendi:2015hoa,Cao:2015cza,Blake:2013bqa,Davison:2013jba,Davison:2013txa,
Zhang:2015nwy,Hu:2015dnl,Do:2016abo,Amoretti:2014zha,Hendi:2015pda}.

In this paper, we will focus on the dRGT massive gravity, and first find out the generalized Vaidya-AdS solutions in this massive gravity by directly solving the gravitational equations. It is easily seen that these Vaidya-AdS solutions are consistent with the results in the previous work in the static spacetime~\cite{Cai:2014znn}. In addition, we also investigate the thermodynamics of these Vaidya-AdS solutions by using the generalized Misner-Sharp energy in the (3+1)-dimensional dRGT massive gravity and unified first law~\cite{Hu:2015xva,Hayward}. Besides the discovery of the thermodynamical first law of these Vaidya-AdS solutions, we also find that the massive gravity is in its thermodynamic equilibrium state, which is consistent with the result considered in the FRW universe~\cite{Hu:2015xva}. Finally, we investigate the holographic thermalization of the boundary field theory which can be holographically modeled by a massive shell collapsing into the Vaidya-AdS black brane. In particular, the thermalization is captured by the equal-time two-point correlation functions, which can be mapped to the bulk by computing the length of the geodesics between these separated points \cite{Balasubramanian1,Balasubramanian2}. Our numerical results show that the graviton mass parameter in dRGT massive gravity can increase the holographic thermalization process, which indicates that the inhomogeneity of the boundary field theory will render the thermalization faster.

This paper is organized as follows. In section II, we find out the generalized Vaidya-AdS solutions in the (3+1)-dimensional dRGT massive gravity;  We adopt the Misner-Sharp energy and unified first law to investigate the thermodynamics of these generalized Vaidya-AdS solutions in section III; In section IV, we investigate the effect of the graviton mass parameter on the holographic thermalization process; Finally, we draw the conclusions and discussions in section V.

\section{Generalized Vaidya-AdS solutions in the 4-dimensional dRGT massive gravity}
Usually, the action of the dRGT massive gravity in an $(n+2)$-dimensional spacetime with the cosmological constant $\Lambda=-\frac{(n+1)n}{2l^2}$ reads~\cite{Vegh:2013sk,Cai:2014znn}
\begin{equation}
\label{actionmassive}
S =\frac{1}{16\pi G}\int d^{n+2}x \sqrt{-g} \left[ R +\frac{n(n+1)}{l^2} +m^2 \sum^4_i c_i {\cal U}_i (g,f)\right],
\end{equation}
where $l$ is the radius of the AdS spacetime, $m^2$ is the graviton mass parameter, and  $f$ is a fixed symmetric tensor, which is usually called the reference metric,
$c_i$ are constants,  and ${\cal U}_i$ are symmetric polynomials of the eigenvalues of the $(n+2)\times (n+2)$ matrix ${\cal K}^{\mu}_{\ \nu} \equiv \sqrt {g^{\mu\alpha}f_{\alpha\nu}}$:
\begin{eqnarray}
\label{eq2}
&& {\cal U}_1= [{\cal K}], \nonumber \\
&& {\cal U}_2=  [{\cal K}]^2 -[{\cal K}^2], \nonumber \\
&& {\cal U}_3= [{\cal K}]^3 - 3[{\cal K}][{\cal K}^2]+ 2[{\cal K}^3], \nonumber \\
&& {\cal U}_4= [{\cal K}]^4- 6[{\cal K}^2][{\cal K}]^2 + 8[{\cal K}^3][{\cal K}]+3[{\cal K}^2]^2 -6[{\cal K}^4].
\end{eqnarray}
The square root in ${\cal K}$ means $(\sqrt{A})^{\mu}_{\ \nu}(\sqrt{A})^{\nu}_{\ \lambda}=A^{\mu}_{\ \lambda}$ and $[{\cal K}]=K^{\mu}_{\ \mu}=\sqrt {g^{\mu\alpha}f_{\alpha\mu}}$ (to extract the roots of the components one by one and then to make summation).  The equations  of motion turns out to be
\begin{eqnarray}
R_{\mu\nu}-\frac{1}{2}Rg_{\mu\nu}-\frac{n(n+1)}{2l^2} g_{\mu\nu}+m^2 \chi_{\mu\nu}&=&8\pi G T_{\mu \nu },~~\label{Equation}
\end{eqnarray}
where
\begin{eqnarray}
&& \chi_{\mu\nu}=-\frac{c_1}{2}({\cal U}_1g_{\mu\nu}-{\cal K}_{\mu\nu})-\frac{c_2}{2}({\cal U}_2g_{\mu\nu}-2{\cal U}_1{\cal K}_{\mu\nu}+2{\cal K}^2_{\mu\nu})
-\frac{c_3}{2}({\cal U}_3g_{\mu\nu}-3{\cal U}_2{\cal K}_{\mu\nu}\nonumber \\
&&~~~~~~~~~ +6{\cal U}_1{\cal K}^2_{\mu\nu}-6{\cal K}^3_{\mu\nu})
-\frac{c_4}{2}({\cal U}_4g_{\mu\nu}-4{\cal U}_3{\cal K}_{\mu\nu}+12{\cal U}_2{\cal K}^2_{\mu\nu}-24{\cal U}_1{\cal K}^3_{\mu\nu}+24{\cal K}^4_{\mu\nu}).
\end{eqnarray}

In this section, we mainly focus on obtaining the generalized Vaidya-AdS solutions with a two dimensional maximally symmetric inner space in the (3+1)-dimensional dRGT massive gravity. Therefore, the general metric ansatz can be
  \be
ds^{2}=-f(v,r)dv^2+2 dv dr+r^{2}\gamma _{ij}dx^{i}dx^{j},
\label{metricn1}
\en
where $\gamma _{ij}$ is the metric on a two-dimensional
constant curvature space ${\cal N}$ with its sectional curvature
$k=\pm 1,0$, and the two-dimensional spacetime ${\cal T}$ spanned by the coordinates $(v,r)$ possesses the metric as $h_{ab}$. The energy momentum tensor for the radiation matter in the spacetime
is given by $T_{ab}=\mu l_al_b$, where $\mu$ is the energy density and $l_a=(1,0,0,0)$ in coordinates $(v,r,x^i)$. In our case, we can also take the reference metric as the following in \cite{Vegh:2013sk,Cai:2014znn}
\begin{equation}
\label{reference}
f_{\mu\nu} = {\rm diag}(0,0, c_0^2 \gamma_{ij} ),
\end{equation}
with $c_0$ being a positive constant. Thus the symmetric polynomials become
\begin{equation}
{\cal U}_1= \frac{2c_0}{r}, ~~{\cal U}_2= \frac{2c_0^2}{r^2},~~ {\cal U}_3= 0,~~ {\cal U}_4= 0,
\end{equation}
Therefore, the field equation in (\ref{Equation}) can be simplified in our case as
\begin{equation}
\mathcal{G}_{\mu\nu} \equiv R_{\mu\nu}-\frac{1}{2}Rg_{\mu\nu}+\Lambda g_{\mu\nu}+m^2 \chi_{\mu\nu}=8\pi G T_{\mu\nu}, \label{Eqn1}
\end{equation}
where $\Lambda=-\frac{3}{l^2}$, and
\begin{eqnarray}
&& \chi_{\mu\nu}=-\frac{c_1}{2}({\cal U}_1g_{\mu\nu}-{\cal K}_{\mu\nu})-\frac{c_2}{2}({\cal U}_2g_{\mu\nu}-2{\cal U}_1{\cal K}_{\mu\nu}+2{\cal K}^2_{\mu\nu}).
\end{eqnarray}
For the metric ansatz in (\ref{metricn1}), the components of the field equation (\ref{Eqn1}) can be explicitly expressed as
\begin{eqnarray}
&&\mathcal{G}_{v}^{v}=\mathcal{G}_{r}^{r}=\frac{-k+\Lambda r^2-c_1c_0m^2r-c_2c_{0}^2m^2+f+rf'}{r^2}=0, \label{MainEq}\\
&&\mathcal{G}^{i}_{j}=\delta^{i}_{j}\times \frac{2r\Lambda -c_1c_0m^2+2f'+rf''}{2r}=0,\\
&&\mathcal{G}_{r}^{v}=\frac{-\dot{f}}{r}=8\pi G \mu,\label{EqST}\\
&&\mathcal{G}_{v}^{r}=0,
\end{eqnarray}
where a prime/overdot denotes the derivative with respect to $r/v$. Note that, the components $\mathcal{G}^{i}_{j}$ are not independent, since it can be found that they can be
linearly expressed in terms of $\mathcal{G}_{v}^{v}$ and $\partial_r{\mathcal{G}_{v}^{v}}$ as $\mathcal{G}^{i}_{j}=\delta^{i}_{j}[\mathcal{G}_{v}^{v}+r\partial_r{\mathcal{G}_{v}^{v}}/2]$, and hence $\mathcal{G}^{i}_{j}=0$ do not yield independent equations.

Therefore, from the above equation in (\ref{MainEq}) and (\ref{EqST}), we can easily obtain the generalized Vaidya-AdS solutions in (3+1)-dimensional dRGT massive gravity
\begin{eqnarray}
&&f(v,r)=k+\frac{r^2}{l^2}-\frac{M(v)}{r}+\frac{c_0c_1m^2}{2}r+c_0^2c_2m^2,\nonumber\\
&&\mu=\frac{\dot{M}(v)}{8\pi G r^2}. \label{Solution}
\end{eqnarray}
Note that, our solutions (\ref{Solution}) can be consistent with those in some previous work like \cite{Cai:2014znn}. Since if $M(v)$ is independent of $v$, i.e. a constant, and hence $f(v,r)$ can be written as $f(r)$, then after making the transformation in the metric ansatz (\ref{metricn1})
\begin{eqnarray}
dv=dt+\frac{1}{f(r)}dr,
\end{eqnarray}
we will easily find that the solutions (\ref{Solution}) are just the same as the static solutions in (3+1)-dimensional spacetime found in \cite{Cai:2014znn} .

\section{Thermodynamics of the generalized Vaidya-AdS solutions}
In this section, we will investigate the thermodynamics of the above generalized Vaidya-AdS solutions by using the unified first law and Misner-Sharp energy.
According to the unified first law \cite{Hayward}, similar to
the case of Einstein gravity, one can usually cast the
equations of gravitational field (\ref{Equation}) in a (3+1)-dimensional spacetime with a two dimensional maximally symmetric inner space into the form
\begin{equation}
dE_{\rm eff}=A\Psi_a dx^a +WdV, \label{2eq5}
\end{equation}
where $A=V_{k}r^2$ and $V =V_{k}r^3/3$ are
area and volume of the $2$-dimensional space with radius $r$. $W$ is called work density defined
as $W= -h^{ab}T_{ab}/2$, while $\Psi_a$ is the energy supply vector, $\Psi_a=T_a^{\ b} \partial _b r +W \partial_a r$. In addition, $T_{ab}$ is the projection of the four-dimensional stress energy $T_{\mu\nu}$ of matter into $h_{ab}$, and $E_{eff}$ is signed as the generalized Misner-Sharp mass in the modified gravity if it exists.

Note that, we have obtained the generalized Misner-Sharp mass in 4-dimensional dRGT massive gravity in \cite{Hu:2015xva}
\begin{eqnarray}
\label{2eq12}
E_{\rm eff}&=&\frac{V_kr}{8\pi G}\left[\left(k-h^{ab}\partial_a r\partial_b r\right)- \frac{\Lambda r^2}{3} + \frac{(c_1c_0m^2r+2c_2c_0^2m^2)}{2}\right].\label{MSCOMG}
\end{eqnarray}%
Therefore, after substituting the explicit forms of the generalized Vaidya-AdS solutions in (\ref{Solution}), we can obtain the following quantities
\begin{eqnarray}
&&E_{\rm eff}=\frac{V_k M(v)}{8\pi G},~~ \Psi_a=\mu l_a,~~W=0,
\end{eqnarray}
From which, we can easily check that the unified first law (\ref{2eq5}) is indeed satisfied for our solutions (\ref{Solution}), and the generalized Misner-Sharp mass indeed exists in the 4-dimensional dRGT massive gravity. Note that, the generalized Misner-Sharp mass does not always exist for the modified gravity, i.e. the $f(R)$ gravity in ~\cite{Cai:2009qf,Zhang:2014goa}.

Next, we will investigate the thermodynamics of the generalized Vaidya-AdS solutions (\ref{Solution}) on the apparent horizon $r_A$, where $r_A$ is defined as the trapped surface $h^{ab}\partial_a{r}\partial_b{r}=0$. In our case, we can easily obtain the location of the apparent horizon $r_A$ subjected to the equation $f(v,r)=0$. In addition, on the apparent horizon, the energy crossing the
apparent horizon within time interval $dv$ is~\cite{CCHK,Cai-Kim,Cai:2008ys,Hu:2015xva}
\begin{equation}
\delta Q=dE_{\rm eff}|_{r_{A}}=A\Psi_adx^a|_{r=r_A} = A\Psi_v dv=\frac{V_k \dot{M}(v)}{8\pi G}dv=-\frac{V_k \dot{f}(r_A)r_A}{8\pi G}dv.
\end{equation}
On the other hand, the temperature of generalized Vaidya-AdS solutions can be considered as $T=\frac{\kappa}{2\pi}$. Here the surface gravity $\kappa$ defined on the apparent horizon is $\kappa=D_aD^ar=\frac{1}{2\sqrt{-h}}\frac{\partial}{\partial x^{\mu}}(\sqrt{-h}h^{\mu\nu}\partial_v r)=f'(r_A)/2$~\cite{Hayward,CCHK,Cai-Kim,Cai:2008ys}, and $D_a$ is the covariant derivative associated with the metric $h_{ab}$. In addition, the entropy of generalized Vaidya-AdS solutions is $S=\frac{A}{4G}=\frac{V_kr_A^2}{4G}$~\cite{Cai:2014znn}. Therefore, we can obtain
\begin{equation}
TdS=\frac{\kappa}{2\pi}dS=\frac{V_k}{8\pi G}f'(r_A)\dot{r}_Ar_Adv.
\end{equation}
Note that, from the equation of the apparent horizon $f(r_A,v)=0$, we can deduce a simple relation $f'(r_A)\dot{r}_A=-\dot{f}(r_A)$. Obviously, after using this simple relation, we can easily find that the usual Clausius relation $\delta Q= TdS$ holds on the apparent horizon for the generalized Vaidya-AdS solutions (\ref{Solution}). Therefore, the unified first law in (\ref{2eq5}) on the apparent can be rewritten as
\begin{equation}
dE_{\rm eff}=TdS +WdV, \label{FirstLaw}
\end{equation}
which is just the first law of thermodynamics for the generalized Vaidya-AdS solutions. In addition, according to~\cite{Eling:2006aw}, the usual Clausius relation $\delta Q= TdS$ held in our case also indicates that the dRGT massive gravity is an equilibrium state. It should be emphasized that the usual Clausius relation $\delta Q= TdS$ does not always hold on the apparent horizon. For example, the usual Clausius relation does not hold for the $f(R)$ gravity, which can be the effect from the nonequilibrium thermodynamics of space-time~\cite{aka,Cai:2009qf,Eling:2006aw}. It should also be pointed out that the usual Clausius relation held or not held only depends on the explicit theory of gravity, i.e. like the case for the formula of entropy $S(A)$ independent of the explicit solutions. Indeed, the usual Clausius relation $\delta Q= TdS$ held in our case is consistent with the investigation in \cite{Hu:2015xva} by taking the FRW universe into account.

\section{Holographic thermalization in massive gravity}
In this section, we will focus on the holographic thermalization in the dRGT massive gravity with the $4$-dimensional bulk spacetime, in which the dual gravitational solution is investigated under the above Vaidya-AdS solution with $k=0$,
\be
ds^{2}=\frac{1}{z^2}[-H(v,z)dv^2-2 dv dz+dx_{i}^2].
\label{metricn2}
\en
where $i=1,2$ representing the codimension 2 space, $z=\frac{l}{r}$ and $H(v,z)=1-M(v)z^3+\frac{c_0c_1m^2}{2}z+c_0^2c_2m^2z^2$. $M(v)$ can be related to the mass of a collapsing black brane, which is usually set as a smooth function
\begin{equation}
M(v) = M \left( 1 + \tanh \frac{v}{v_0} \right),\label{M}
\end{equation}
in which $v_0$ can represent the finite thickness of a null shell and $M$ is a constant. From Eq.(\ref{M}), we can see that
in the limit $v\rightarrow-\infty$, the mass vanishes and the background in Eq.(\ref{metricn2}) thus  corresponds to a pure  AdS space with corrections in graviton mass $m^2$; In the limit $v\rightarrow \infty$, the mass turns out to be a constant and the background represents a static AdS black brane in the massive gravity.

Holographic thermalization has been studied from various aspects, such as colliding shock waves, gravitational collapsing, holographic (quantum)quenches, holographic entanglement entropy and etc., see reviews \cite{CasalderreySolana:2011us,DeWolfe:2013cua}. In this section we will focus on the gravitational collapsing aspect and adopt the two-point correlation function at equal time to explore the thermalization process in the dual field theory, and then see how the graviton mass parameter affects this process. According to the AdS/CFT correspondence, the on-shell equal time two-point correlation function can be holographically approximated as \cite{Balasubramanian2}
\begin{equation}
\langle {\cal{O}} (t_0,x_i) {\cal{O}}(t_0, x_i^{\prime})\rangle  \approx
e^{-\Delta { L}} ,\label{llll}
\end{equation}
in which the conformal dimension $\Delta$ of scalar operator $\cal{O}$
is assumed to be large enough, and
$L$ is the bulk geodesic length between the points $(t_0,
x_i)$ and $(t_0, x_i^{\prime})$ on the AdS boundary.  Usually the geodesic length $L$ is divergent
 due to the UV contribution from the AdS boundary, therefore, one should eliminate the  divergence by adding a counter-term in order to obtain the renormalized geodesic length which is $\delta L=L+2\ln z_0$ \cite{Balasubramanian1, Balasubramanian2}, where $2\ln z_0$ comes from the contribution of a pure AdS boundary, and $z_0$ is a UV cut-off satisfying the boundary conditions
\begin{equation}\label{regularization}
z\left(\frac{d}{2}\right)=z_0, ~~~v\left(\frac{d}{2}\right)=t_0,
\end{equation}
in which $d$ is the  spatial separation between the two points lies entirely in the $x_1$ direction while $t_0$ is the time of the thermalization probe moving from the shell to the boundary, which will be recognized as thermalization time later.

In the following context, we would like to rename $x_1$ as $x$ for simplicity and adopt it to parameterize the geodesic. Considering the parity symmetry in $x$ direction $x\to -x$, the proper length of the geodesic is given by
 \begin{eqnarray}
L=2 \int_0^{\frac{d}{2}} dx \frac{\sqrt{\Pi}}{z} ,\label{false}
\end{eqnarray}
with
$
\Pi=1-2z'(x)v'(x) - H(v,z) v'(x)^2
$.
In order to compute the length of the geodesic, we need to minimize the length \eqref{false} and solve the two coupled Euler-Lagrangian equations of motion for $z(x)$  and  $v(x)$ respectively. We reach
\begin{eqnarray} \label{gequation}
&&z(x)\sqrt{\Pi}\partial_x \left(\frac{z'(x)+H(v,z)v'(x)}{z(x)\sqrt{\Pi}}\right)=\frac{1}{2}\frac{\partial H(v,z) }{\partial v(x)} v'^2(x),\nonumber\\
&&z(x)\sqrt{\Pi}\partial_x \left(\frac{v'(x)}{z(x)\sqrt{\Pi}}\right)=\frac{1}{2}\frac{\partial H(v,z) }{\partial z(x)} v'^2(x)+ \frac{\Pi}{ z(x)}.
\end{eqnarray}
In order to solve the above equations, we impose the following boundary conditions of $z(x)$ and $v(x)$ at $x=0$,
\begin{equation}\label{initial}
z(0)=z_*, ~~ v(0)=v_* ,~~ v'(0) =z'(0) = 0.
\end{equation}
where $z'(0)=v'(0)=0$ can be deduced from the symmetry of the geodesic.

One can further simplify the EoMs \eqref{gequation} by virtue of the `Hamiltonian' methods \cite{Ryu:2006ef}. If regarding the integrand $\sqrt\Pi/z$ in \eqref{false} as the `Lagrangian' $\mathcal{L}$, we can define the `Hamiltonian' $\mathcal{H}$ as
\begin{eqnarray}\label{ham}
\mathcal{H}=\mathcal{L}\left(v(x),z(x)\right)-v'(x)\frac{\partial\mathcal{L}}{\partial v'(x)}-z'(x)\frac{\partial\mathcal{L}}{\partial z'(x)}=\frac{1}{z(x)\sqrt{\Pi}}.
\end{eqnarray}
 In this case, $x$ actually plays the role as the time coordinate in classical mechanics while $\mathcal{H}$ as the Hamiltonian. Note that the Hamiltonian $\mathcal{H}$ now does not explicitly contain the variable $x$, therefore, it is conservative w.r.t. the coordinate $x$, i.e., $\mathcal{H}=\frac{1}{z\sqrt{\Pi}}\equiv\text{const. in} ~x$. Hence at $x=0$, it is easy to get that
\be\label{pi}
\frac{1}{z\sqrt\Pi}=\frac{1}{z_*}\Rightarrow \sqrt\Pi=\frac{z_*}{z}.
\en
Therefore, substituting \eqref{pi} into the EoMs \eqref{gequation} we can substantially simplify the equations as
\begin{eqnarray}\label{vzeq}
v''(x)-\frac12\frac{\partial H(v,z)}{\partial z(x)}v'(x)^2-\frac{\Pi}{z(x)}&=&0,\\
z''(x)+H'(x)v'(x)+\frac12H(v,z)\frac{\partial H(v,z)}{\partial z(x)}v'(x)^2-\frac12\frac{\partial H(v,z)}{\partial v(x)}v'(x)^2+\frac{H(v,z)\Pi}{z}&=&0.
\end{eqnarray}
So finally we obtain the second order ordinary differential equations for $v(x)$ and $z(x)$, which can be solved numerically if we impose the boundary conditions in \eqref{regularization} and \eqref{initial}.

\begin{figure}[h]
\centering
\subfigure[$v_{\star}=-0.888, m^2=1$]{\includegraphics[trim=0cm 1.3cm 12cm 22cm, clip=true, scale=0.58]{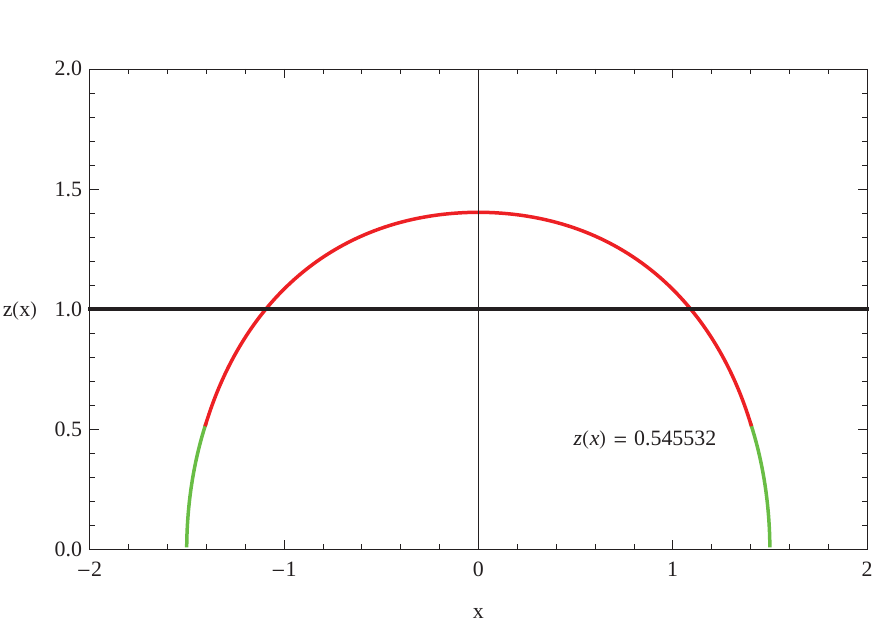} }
\subfigure[$v_{\star}=-0.888, m^2=0.5$]{\includegraphics[trim=0cm 1.cm 0cm 1.cm, clip=true, scale=0.175]{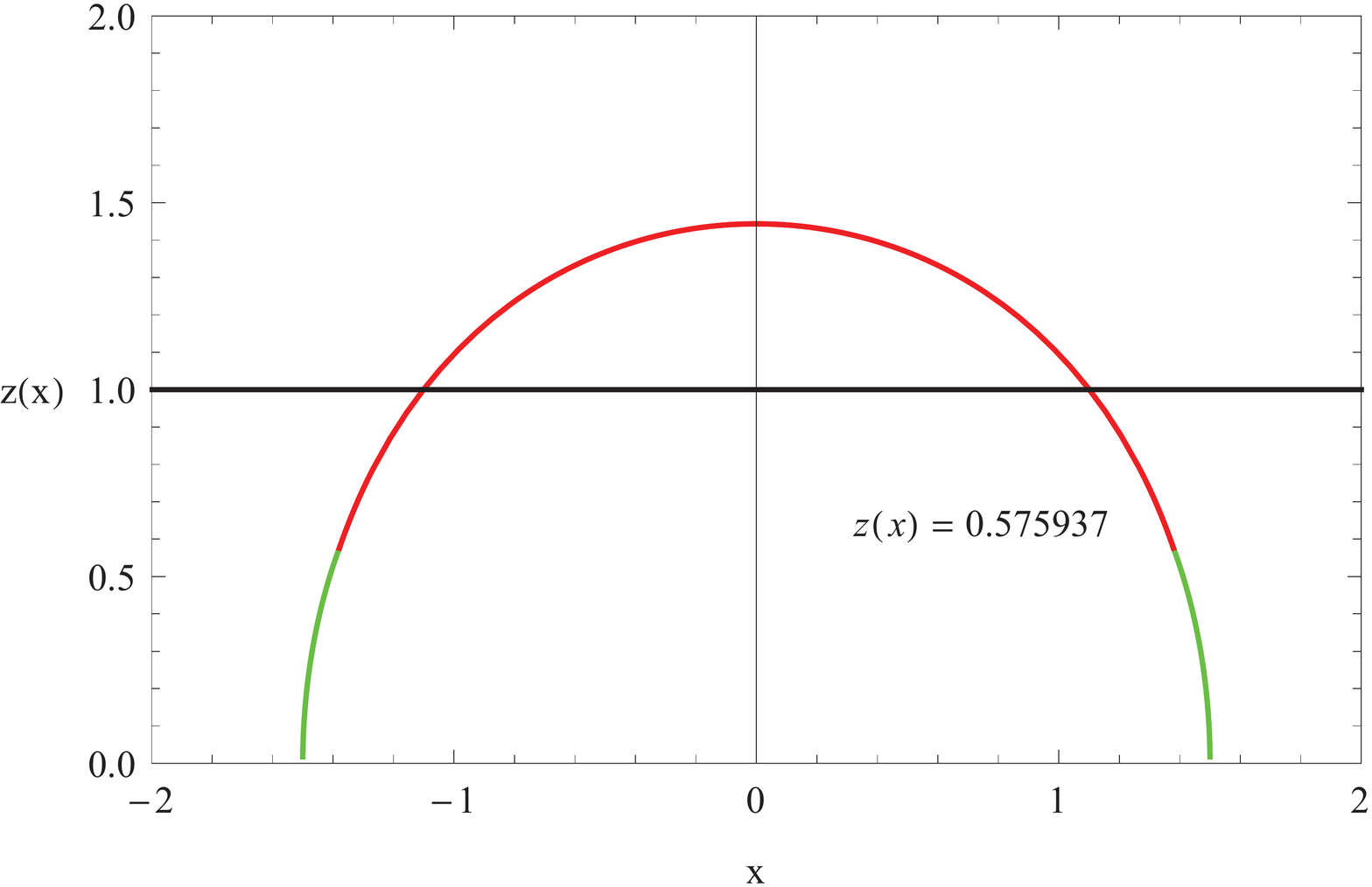} }
\subfigure[$v_{\star}=-0.888, m^2=0.0001$]{\includegraphics[trim=0cm 1.3cm 12.cm 22cm, clip=true,scale=0.58]{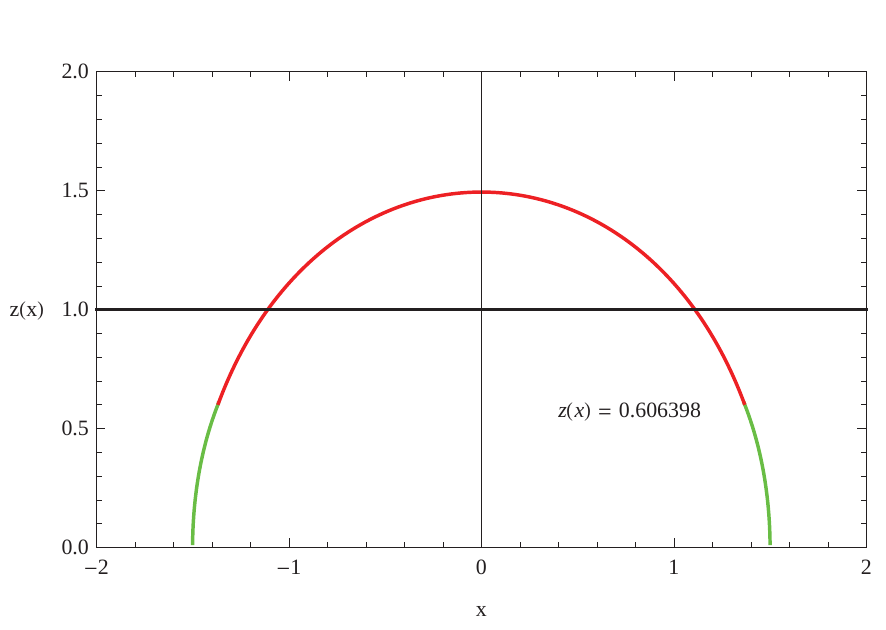}}\\
\subfigure[$v_{\star}=-0.555, m^2=1$]{\includegraphics[trim=0cm 1.3cm 12cm 22cm, clip=true,scale=0.58]{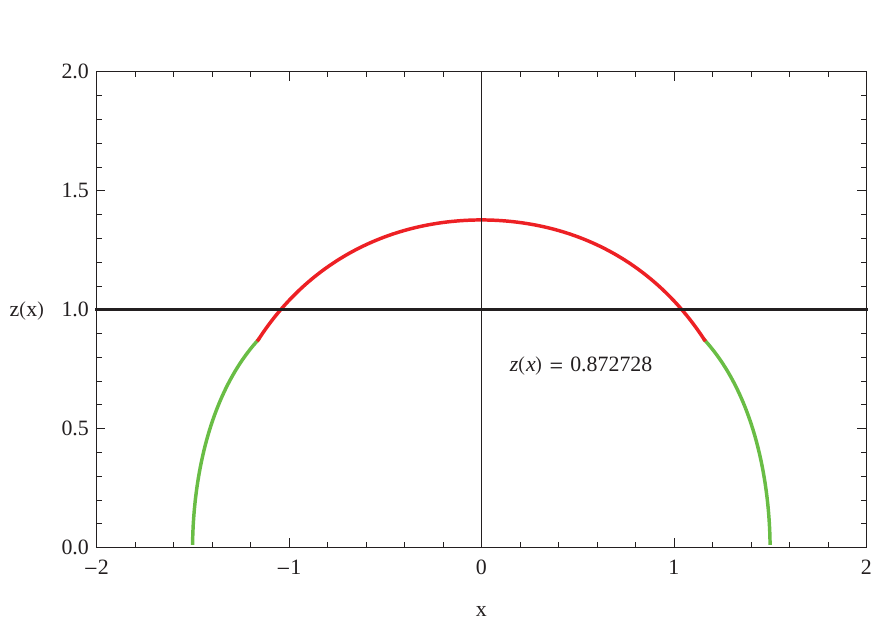}}
\subfigure[$v_{\star}=-0.555,  m^2=0.5$]{\includegraphics[trim=0cm 1.cm 0cm 1.cm, clip=true, scale=0.175]{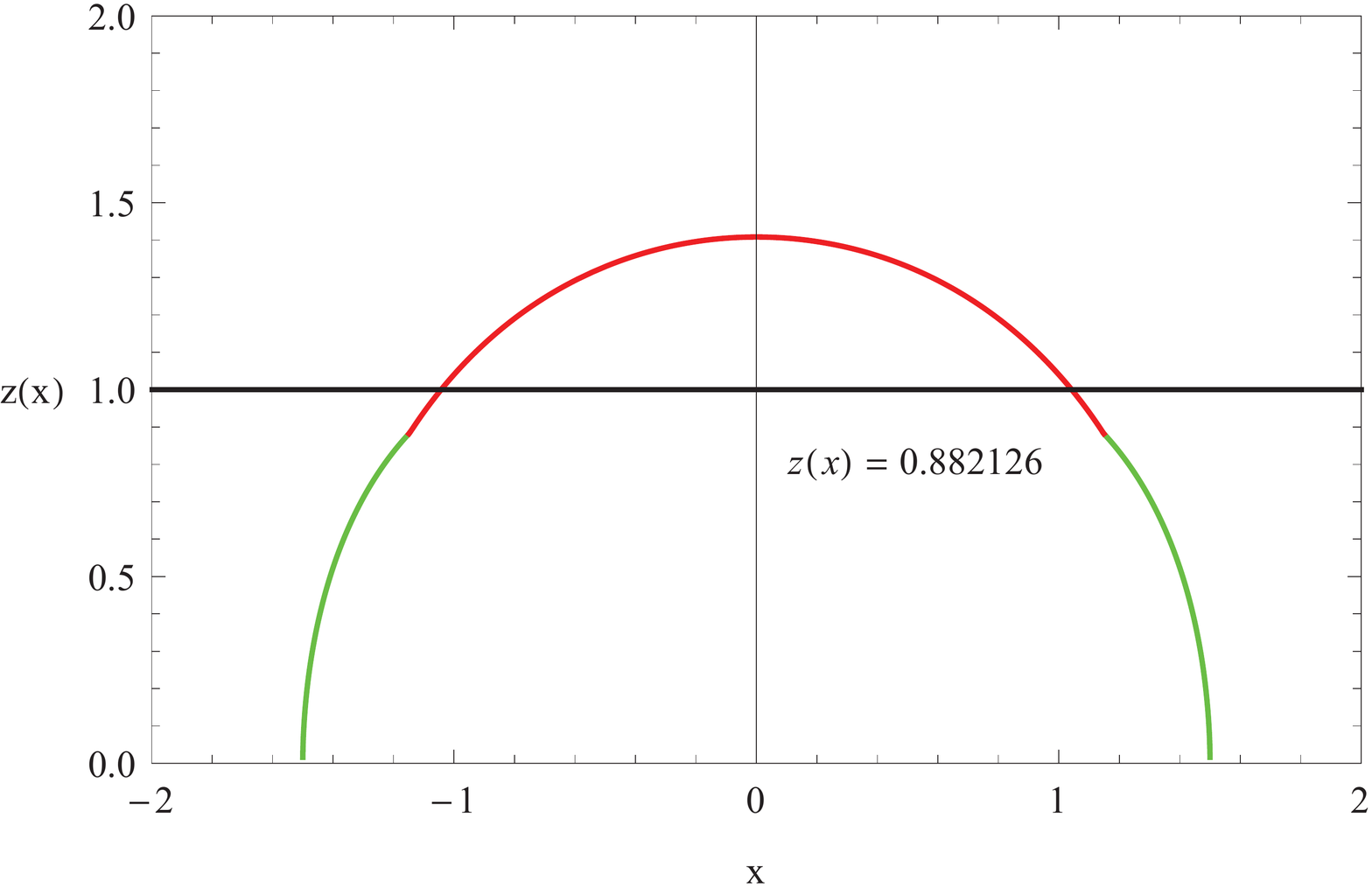} }
\subfigure[$v_{\star}=-0.555, m^2=0.0001$]{\includegraphics[trim=0cm 1.3cm 12cm 22cm, clip=true,scale=0.58]{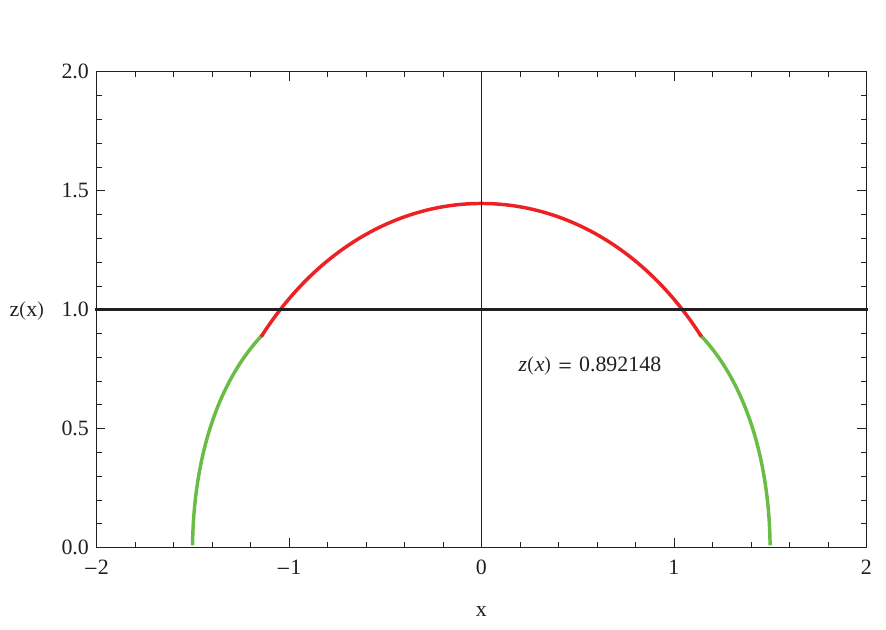}}\\
\subfigure[$v_{\star}=-0.222, m^2=1$]{\includegraphics[trim=0cm 1.cm 0cm 1.cm, clip=true, scale=0.175]{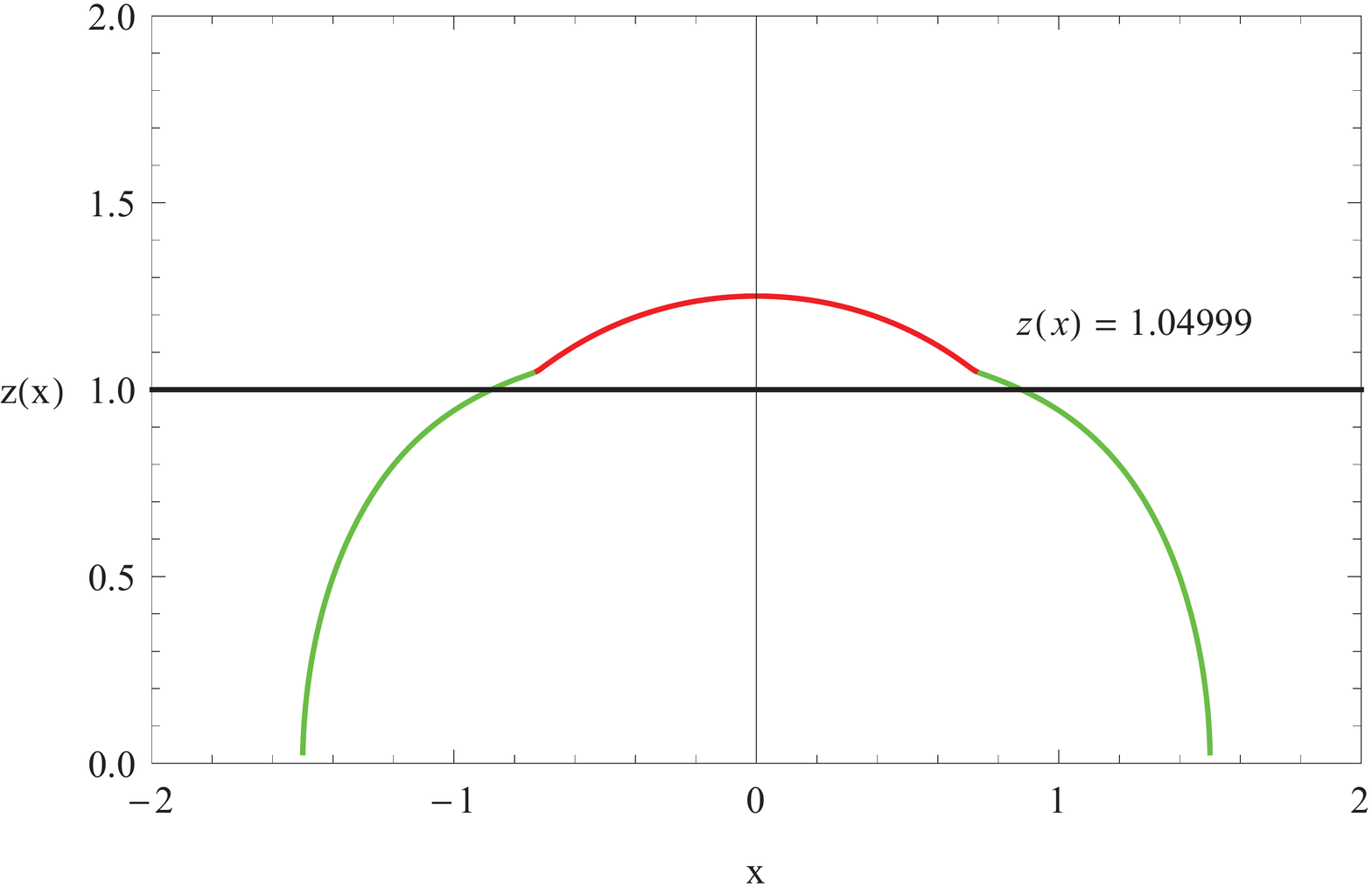}}
\subfigure[$v_{\star}=-0.222, m^2=0.5$]{\includegraphics[trim=0cm 1.cm 0cm 1.cm, clip=true, scale=0.175]{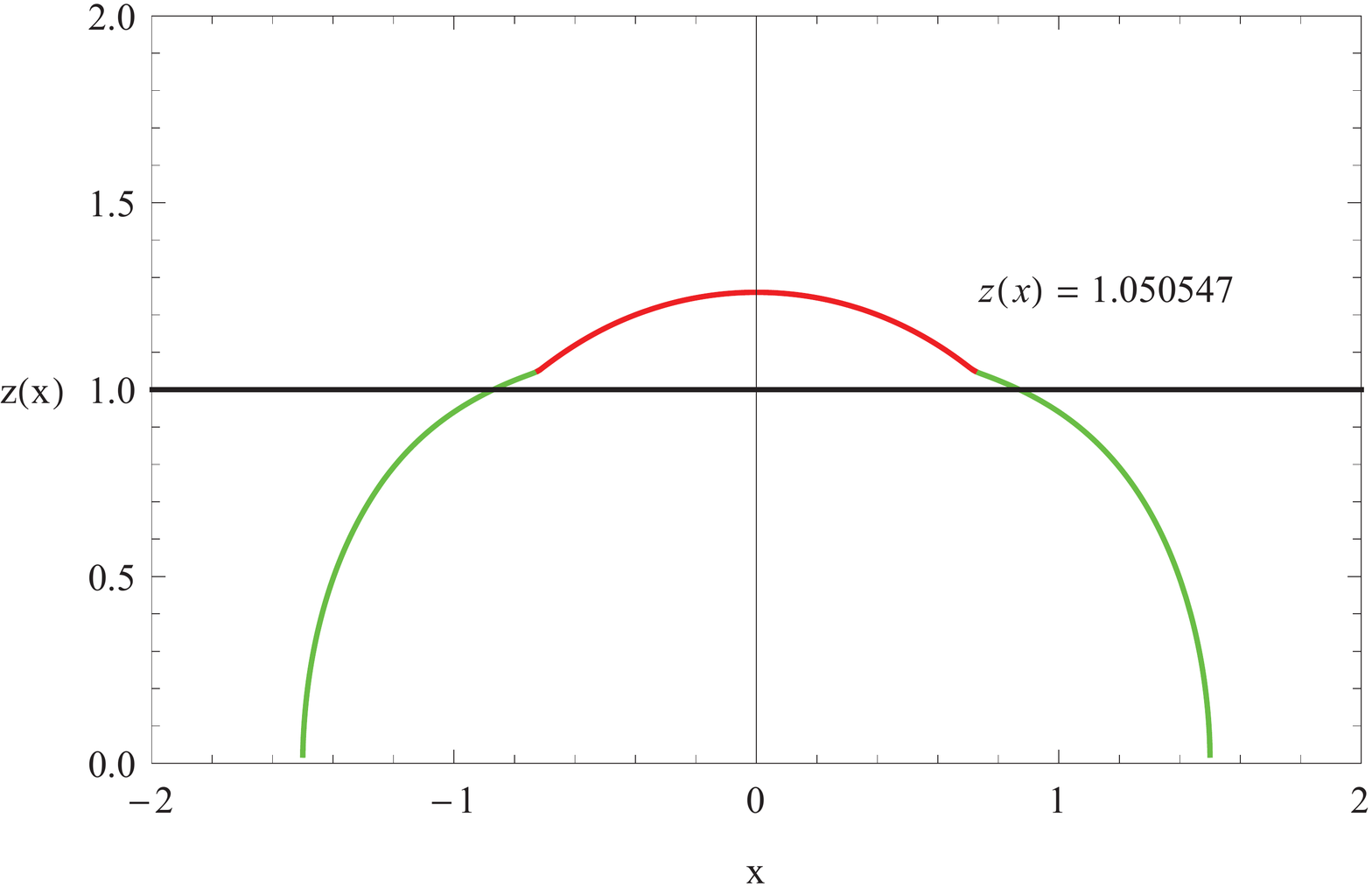} }
\subfigure[$v_{\star}=-0.222, m^2=0.0001$]{\includegraphics[trim=0cm 1.cm 0cm 1.cm, clip=true, scale=0.175]{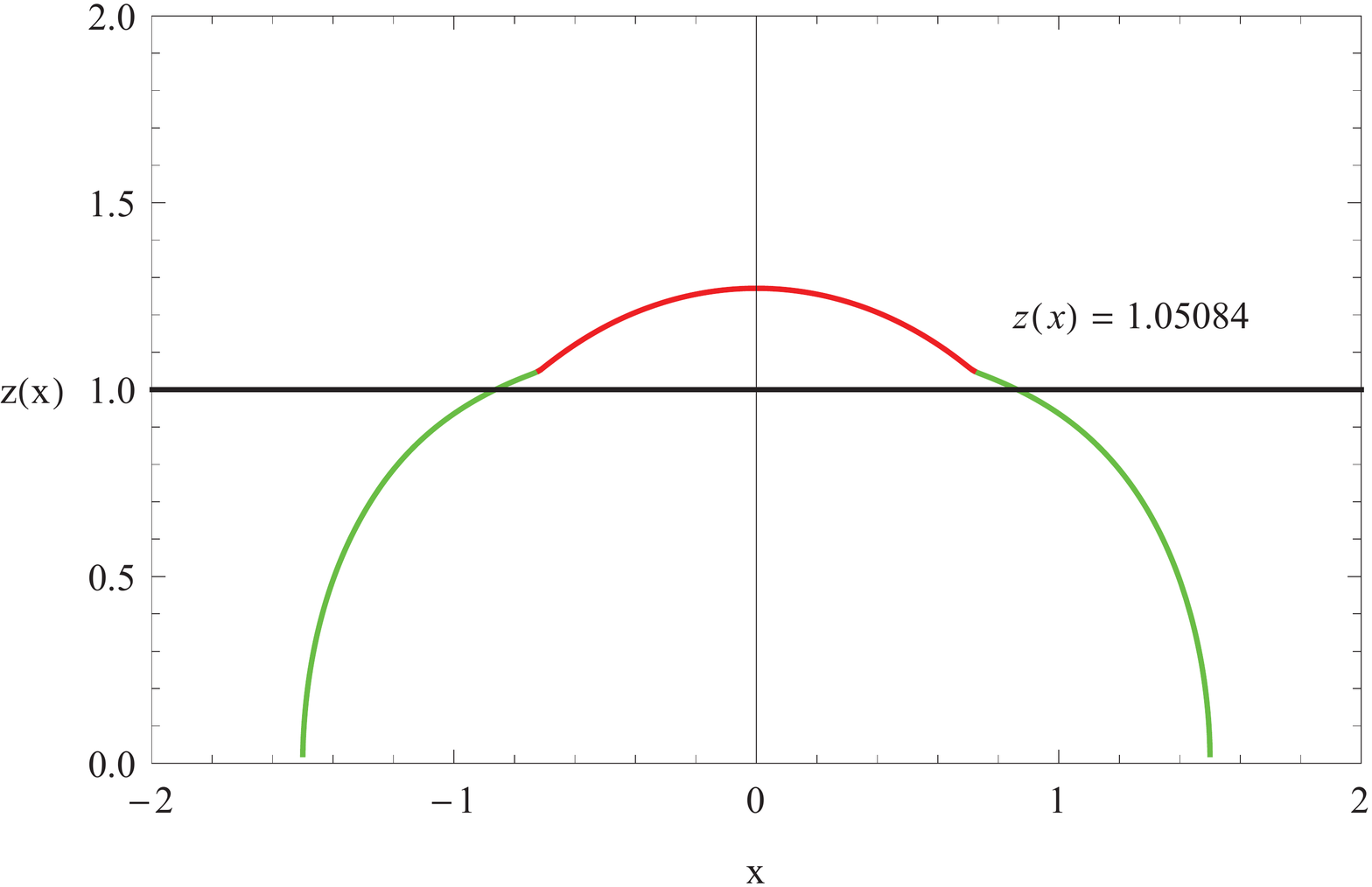} }
 \caption{\small Motion profile of the geodesics in the background of AdS black brane. The separation of the points in the
boundary field theory  is $d=3$. The black brane horizon is indicated by the horizontal black line and rescaled to $z=1$. The location of the shell is indicated by the junctions between the red line and the green line.} \label{fig1}
\end{figure}

In the following we will present the numerical results of the above differential equations. In the numerics we fixed the parameters $c_0=c_1=1, c_2=-1/2$ in order to render the background thermodynamically stable \cite{Cai:2014znn,Hu:2015dnl}; Besides, we set the shell thickness $v_0=0.01$ and the UV cutoff at $z_0=0.01$, respectively. In addition, we rescaled the horizon location to be $z=1$ in order to more easily compare the positions of the horizon and the shell. In Fig.\ref{fig1} we show the plots of the space-like geodesics for various initial time $v_*$ and the graviton mass $m^2$. The distances between the boundary points are set to $d=3$ for all the plots. The black horizontal lines represent the horizons of the black brane. The locations of the shells are described by the junctions between red and green lines. For a fixed initial time, one can find that the positions of the shells decrease as the graviton mass $m^2$ increases. For a fixed graviton mass $m^2$, the locations of the shells increase as the initial time increase. When the initial time is relatively small, for instance $v_*=-0.888$ and $v_*=-0.555$ in the first two lines of Fig.\ref{fig1}, the shells are outside the horizon which means the shells are still in the process of thermalizing. On the contrary, if the initial time is a little bit bigger such as $v_*=-0.222$ in the bottom line of Fig.\ref{fig1}, the shells have already dropped into the horizon which indicates that the dual field theory in the boundary has already been thermalized and in an equilibrium state.

In Table.\ref{tab1} we present the data of the thermalization times $t_0$ for various graviton masses and initial times. One finds that for a fixed initial time $v_*$, the thermalization time $t_0$ decreases as the graviton mass increases, which indicates that the dual boundary field theory thermalizes and saturates into the equilibrium state faster. As we learned from \cite{Blake:2013owa}, graviton mass is related to the inhomogeneity of the boundary field theory. Therefore, from the Table.\ref{tab1} we can infer that on the boundary field theory, the greater the inhomogeneity is, the faster the system would saturate into the equilibrium state.
\begin{table}[h]
\begin{center}\begin{tabular}{|l|c|c|c|c|c}
 %\MC{3}{c}{\text{caption}}\\[5pt]
 \hline
% & \multicolumn{3}{c||}{MGCDM}   & \multicolumn{3}{c}{$\Lambda$CDM}  \\ \hline
%                             &        MGCDM        &                  &             &      $\Lambda$CDM    &                   & \\ \hline
% \MC{3}{|c|c|}{\ZZ
%{15pt}\hfill\normalsize   \hfill  \hfill\normalsize MGCDM     \hfill\normalsize $\Lambda$CDM  }\\ \hline
% \ZZ{-6pt}{22pt}
                    &$m^2=1$      & $m^2=0.5$  &  $m^2=0.0001$  \\ \hline
$v_{\star}=-0.888$    & $0.492817$   &$0.542195$  & $ 0.597428$    \\ \hline
$v_{\star}=-0.555$   &$0.805171$   &$0.851048$  &$0.901887 $     \\ \hline
$v_{\star}=-0.222$     &$1.0924$  & $1.13325$   &$1.17788 $  \\ \hline
\end{tabular}
\end{center}
\caption{The thermalization time $t_0$ of the geodesic probe for different graviton mass $m^2$ initial time $v_{\star}$. The separation of points on the boundary is $d=3$.}\label{tab1}
\end{table}

\begin{figure}[h]
\centering
\subfigure[$ d=2$]{
\includegraphics[trim=0cm 1.5cm 0cm 1.5cm, clip=true, scale=0.26]{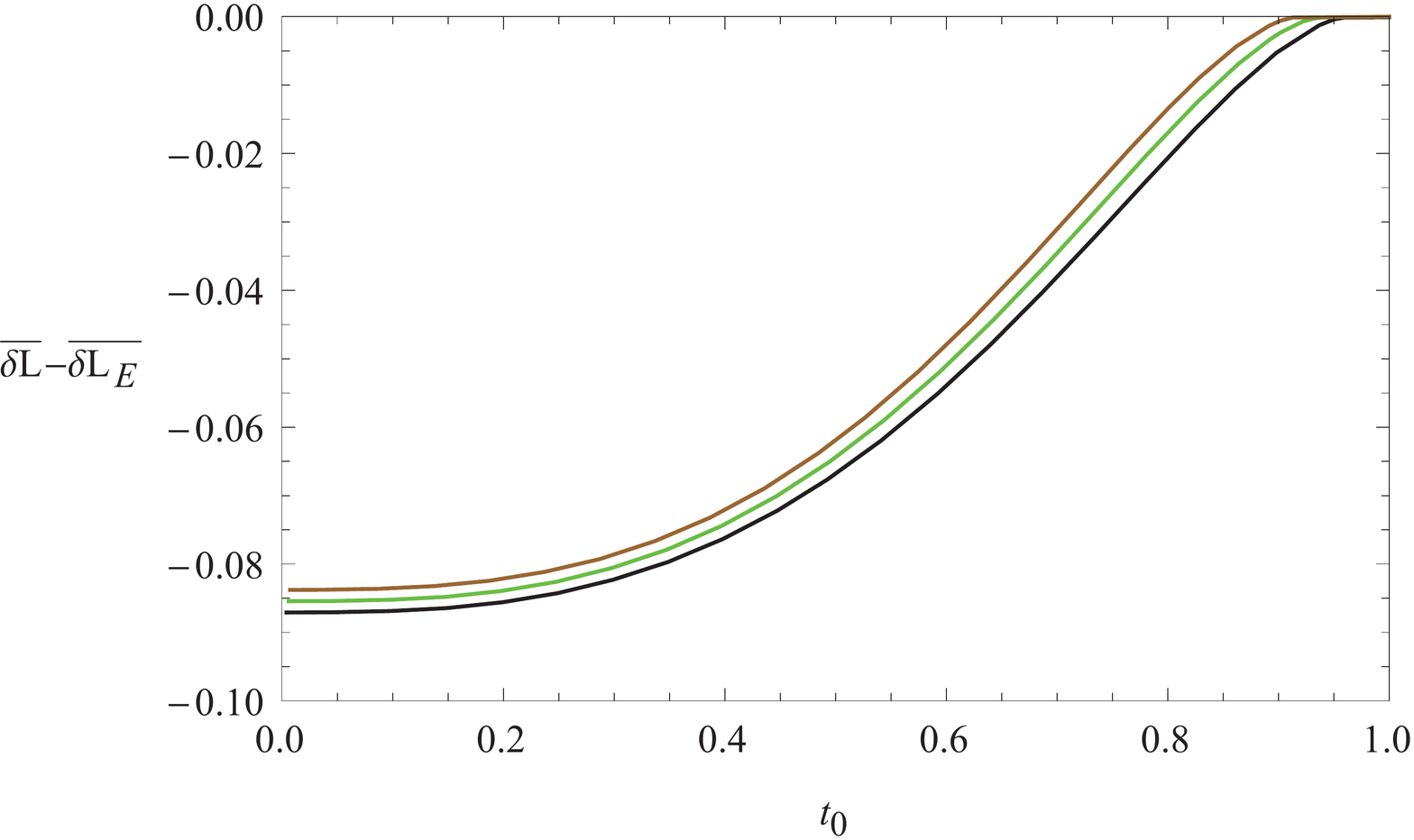}  }
\subfigure[$ d=3$]{
\includegraphics[trim=0cm 1.5cm 0cm 1.5cm, clip=true, scale=0.26]{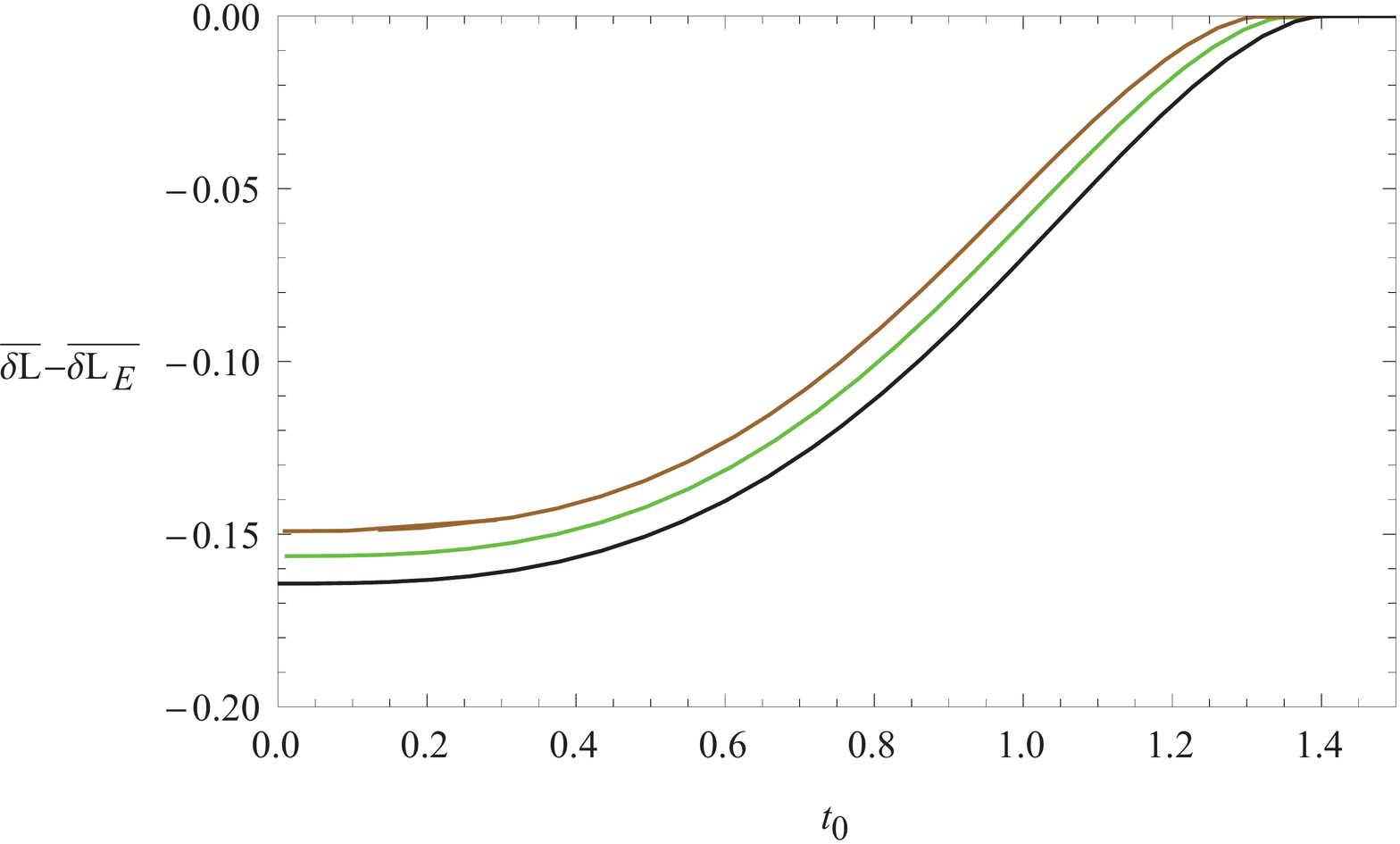}
}
 \caption{\small Thermalization from the renormalized geodesic lengths for various graviton masses with the distances of the separated points $d=2$ (panel $(a)$) and $d=3$ (panel $(b)$).
 The brown, green and black  lines are corresponding to graviton masses  $m^2=1, 0.5, 0.0001$ respectively.}  \label{fig2}
\end{figure}

In Fig.\ref{fig2}, we plot the renormalized geodesic length for different graviton masses and different separation of the points on the boundary. In particular we define a renormalized dimensionless geodesic length as $\overline{\delta L}-\overline{\delta L_E}$, in which $\overline{\delta L}=\delta L/d$ and $\overline{\delta L_E}$ is the value of the geodesic length arriving at equilibrium state. The brown, green and black lines in the Fig.\ref{fig2} correspond to graviton masses $m^2=1, 0.5, 0.0001$ respectively. Therefore, $\overline{\delta L}-\overline{\delta L_E}=0$ indicates that the system saturates an equilibrium thermal state. Hence, it is readily to see that as graviton mass becomes bigger, the corresponding boundary field theory saturates into equilibrium faster, which is consistent with the conclusion drawn from Table.\ref{tab1}. Once again this indicates that greater inhomogeneity in the boundary field theory would make the system faster to enter the equilibrium state. From the panels $(a)$ and $(b)$ in Fig.\ref{fig2}, we can also deduce that different separations of the points on the boundary will not change the conclusions above.

\section{conclusion and discussion}
In this paper, after obtaining the generalized Vaidya-AdS solutions by directly solving the equations of fields in the dRGT massive gravity, we have investigated the thermodynamics of these generalized Vaidya-AdS solutions. Besides the first law of thermodynamics obtained by using the unified first law and generalized Misner-Sharp mass, we also find that the usual Clausius relation $\delta Q= TdS$ holds in our case, which indicates that the dRGT massive gravity is a equilibrium state. This result is consistent with that by taking the FRW universe into account. Moreover, we further investigate the holographic thermalization by considering a massive shell collapsing in the generalized Vaidya-AdS black brane spacetime, while the two-point correlation function at equal time has been chosen as a thermalization probe to investigate the thermalization behavior of the dual field on the boundary. On the other hand, according to the AdS/CFT correspondence, the two-point correlation function at equal time can be related to the length of the space-like geodesics in the bulk between the points $(t_0, x_i)$ and $(t_ 0, x '_ i)$ on the AdS boundary. Therefore, after some numerical calculations, our final result is that the graviton mass parameter can increase the holographic thermalization process.

Note that, some clues have been shown that the effect from the massive graviton in the bulk can be considered as the effect from the lattice in the dual field according to the applications of AdS/CFT in condensed matter, i.e. AdS/CMT. Therefore, an interesting question related to our results is that what is the dual physical meaning of the effect from the massive graviton for the field on the AdS boundary? In addition, another interesting question is about the relationship between the graviton mass and the two-point correlation function at equal time on the AdS boundary. From our results, we can qualitatively find that the graviton mass affects the length of the bulk geodesic, and hence the two-point correlation function at equal time through (\ref{llll}), which finally deduce the thermalization time shorter. Note that, recently there are also some investigations on the inhomogeneous holographic thermalization \cite{Adams:2012pj,Garcia-Garcia:2013rha}, which will be also an interesting issue to be further studied. These investigations may give more insights to the information about the real thermalization process of QGP. On the other hand, it should be pointed out that other choices of probes like Wilson loops or entanglement entropy are also possible if the bulk spacetime is of dimension $D > 3$, i.e., the holographic thermalization process takes place in the space with dimension $D>1$. Moreover, it is well known that in the Einstein gravity the entanglement entropy is the probe that thermalizes later and therefore sets the thermalization time of the field theory, thus an interesting future direction would be to know whether the dRGT massive gravity also follows the same pattern as Einstein gravity, namely that probes of codimension $2$ are the ones that take longer to thermalize.

\section{Acknowledgements}
This work is supported by the National Natural Science Foundation of China (NSFC) (Grant Nos. 11575083, 11565017, 11105004,11405016, 11675140), the Fundamental Research Funds for the Central Universities (Grant No. NS2015073), China Postdoctoral Science Foundation (Grant No. 2016M590138), and the Open Project Program of State Key Laboratory of Theoretical Physics, Institute of Theoretical Physics, Chinese Academy of Sciences, China (Grant  No. Y5KF161CJ1). In addition, Y.P Hu thanks a lot for the support from the Sino-Dutch scholarship programme under the CSC scholarship.

\end{document}